\documentclass[%
 reprint,
nobibnotes,
 amsmath,amssymb,
 aps,
]{revtex4-1}

\usepackage{graphicx}
\usepackage{dcolumn}
\usepackage{bm}
\usepackage{textcomp}
\usepackage{comment}
\usepackage{float}

\begin{document}

\preprint{APS/123-QED}

\title{Accelerating speckle suppression by a degenerate cavity laser with an intracavity phase diffuser}

\author{Simon Mahler$^{1\ddagger}$}
\author{Yaniv Eliezer$^{2\ddagger}$}
\author{Hasan Y{\i}lmaz$^2$}
\author{Asher A. Friesem$^1$}
\author{Nir Davidson$^1$}
\author{Hui Cao$^2$}
\email{hui.cao@yale.edu}

\affiliation{$^1$Department of Physics of Complex Systems, Weizmann Institute of Science, Rehovot 761001, Israel}
\affiliation{$^2$Department of Applied Physics, Yale University, New Haven, Connecticut 06520, USA}

\affiliation{$\ddagger$ These authors contributed equally to this work.}

\date{\today}

\begin{abstract}
Fast speckle suppression is crucial for time-resolved full-field imaging with laser illumination. Here, we introduce a method to accelerate the spatial decoherence of laser emission, achieving speckle suppression in the nanosecond integration time scale. The method relies on the insertion of an intracavity phase diffuser into a degenerate cavity laser to break the frequency degeneracy of transverse modes and broaden the lasing spectrum. The ultrafast decoherence of laser emission results in the reduction of speckle contrast to $3\%$ in less than $1$ nanosecond.
\end{abstract}

                              
                              
\maketitle

\section{Introduction}

Conventional lasers have a high degree of spatial coherence, manifesting coherent artifacts and cross-talk. One prominent example is speckle noise, which is detrimental to laser applications such as imaging, display, material processing, photolithography, optical trapping and more~\citep{Goodman10}. Several techniques have been developed to suppress speckle noise by incoherently integrating many uncorrelated speckle realizations, e.g. by using a moving diffuser or aperture ~\citep{Chellappan10, Lowenthal71, McKechnie75, Saloma90, Kuratomi10, Waller12, Tran:16, akram2016speckle}.
Typically, these methods are effective only at long integration times, i.e. in the order of a millisecond or longer. 

Fast speckle suppression is essential for time-resolved imaging of moving targets or transient phenomena \cite{MermillodBlondin13, Chriki18, Knitter16}. It can be achieved by using multimode lasers with low and tunable spatial coherence on demand~\citep{redding2011spatial, Redding12, Nixon13_2, Redding15, Hokr16, Liew17, Kyungduk19}. The decoherence time of such lasers, critical for fast speckle suppression in short integration times, is determined by the frequency spacing and linewidth of the individual lasing modes, as well as the total width of emission spectrum $\Delta \Omega$~\cite{Cao19}. Let us consider $N$ transverse modes lase simultaneously, and assume that the linewidth of each individual transverse mode is smaller than the typical frequency spacing $\Delta \omega_t$ of neighboring modes. Only when the integration time $\tau$ exceeds $1/\Delta \Omega$, the modal decoherence starts. Once $\tau$ exceeds $1 / \Delta \omega_t$, the $N$ lasing modes become mutually incoherent and the speckle contrast $C$ is reduced to $1/ \sqrt{N}$. Therefore, broadening the laser emission spectrum and increasing the frequency spacing between the transverse modes accelerates speckle suppression, as demonstrated recently with a broad-area semiconductor laser~\citep{Kyungduk19}.

To reach low spatial coherence, a large number of transverse modes must lase simultaneously. This requires the modes to have a similar loss or quality factor, which can be achieved with a degenerate laser cavity (DCL)\cite{Arnaud69}. The DCL self imaging configuration ensures that all transverse modes have an almost identical (degenerate) quality factor. Experimentally It has been shown that $N \approx 320,000$ transverse modes can lase simultaneously and independently in a solid-state DCL~\citep{Nixon13}. However, the transverse modes are also nearly degenerate in frequency, which implies a longer decoherence (integration) time. In the short nanosecond time scale, the longitudinal modes play a critical role in spatial coherence reduction~\citep{Chriki18}. In particular, the spatio-temporal dynamics of a DCL having $M$ longitudinal modes reduces the speckle contrast to $1/\sqrt{M}$. However, since the number of longitudinal modes is typically far less than the number of transverse modes ($M\ll N$), this method yields limited speckle contrast reduction at short time scales.

In this work, we accelerate the spatial decoherence of a DCL by inserting a phase diffuser (random phase plate) into the cavity. The intracavity phase diffuser lifts the frequency degeneracy of transverse modes and broadens the lasing spectrum. Simultaneously, a large number of transverse modes manage to lase because of their high quality factors. The speckle contrast is reduced to  $3 \%$ in less than one nanosecond. The laser output power is reduced by merely $15 \%$ with the intracavity phase diffuser. This work provides a simple and robust method for ultrafast speckle suppression.

\section{Laser cavity configuration}

\begin{figure*}[ht] 
	\centering
	\includegraphics[width=1\linewidth]{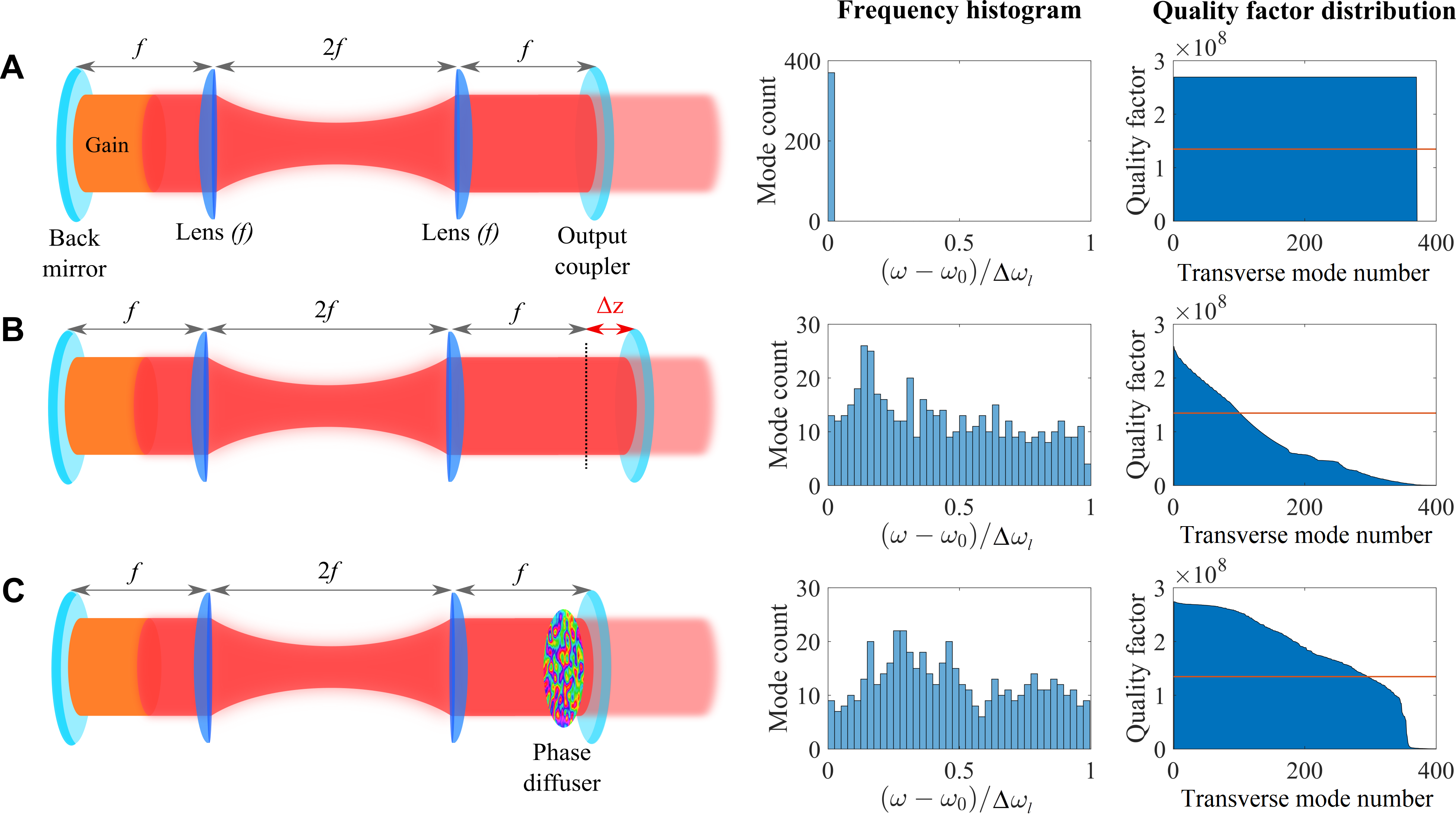}
	\caption{Degenerate cavity laser (DCL) configurations. (A) An Ideal DCL with a total length of $4f$. (B) A misaligned DCL. The output coupler is longitudinally translated by $\Delta z$ along the cavity axis $z$, with a total cavity length of $4f+\Delta z$. (C) A DCL with an intracavity phase diffuser placed next to the output coupler. The left column contains a sketch of each cavity configuration, the middle column shows the histogram of the frequency differences between the transverse modes and the fundamental one within one longitudinal mode group normalized by the free spectral range: $(\omega - \omega_0) / \Delta \omega_l$, and the right column plots the quality factor vs. transverse mode index. The red horizontal line marks half of the maximum quality factor as a reference. The ideal DCL has a large number of high-quality transverse modes, all with the same frequency (both frequency and quality factor degeneracies). The longitudinally-misaligned DCL has many modes with different frequencies but also has a relatively small number of transverse modes with high quality factors (no degeneracies). The DCL with a phase diffuser has a relatively large number of high-quality transverse modes with enhanced frequency differences, enabling ultrafast speckle suppression.}
	\label{fig:1_exp_skecth}
\end{figure*}

Figure~\ref{fig:1_exp_skecth}A is a sketch of our DCL in a self-imaging condition~\citep{Nixon13}. It is comprised of a high-reflectivity flat back mirror, a Nd:YAG gain medium optically pumped by a flash lamp, two spherical lenses of focal lengths $f$ in a $4f$ telescope configuration and an output coupler, more details are given in Methods (1). We calculate the transverse modes (see Methods for details), and plot the histogram of the frequency differences between the $n$-th order transverse mode $\omega_n$ and the fundamental mode $\omega_0$. $\omega_n - \omega_0$ is normalized by the free spectral range (FSR), which is the frequency spacing of longitudinal mode groups $\Delta \omega_l$. The panel in the center indicates that all the transverse modes in a perfect DCL are exactly degenerate in frequency. The quality factor as a function of the transverse mode index in the right plot exhibits a uniform distribution of high quality factors, indicating that all the transverse modes have an exactly identical (degenerate) quality factor. In this ideal case, despite the fact that many transverse modes are expected to lase, the spectral degeneracy slows down the spatial decoherence. Only when the photodetection integration time exceeds the coherence time given by the inverse of spectral linewidth of individual lasing modes, the degenerate modes become mutually incoherent and the speckle contrast decreases.
Note that in practice such an ideal DCL cannot be realized due to the presence of misalignment errors, thermal effects, and optical aberrations~\cite{Arnaud69_2}. Therefore, the transverse lasing modes have slightly different frequencies, which in turn shortens the time of decoherence~\citep{Nixon13}.

In order to accelerate the spatial decoherence, the frequency spacing of the transverse modes has to be increased. Namely, the frequency degeneracy of the modes has to be broken. A conventional method for breaking the frequency degeneracy is detuning the cavity, e.g. translating the output coupler in the longitudinal ($z$) axis of the cavity, which leads to the configuration in Figure~\ref{fig:1_exp_skecth}B. With a sufficient longitudinal displacement $\Delta z$,  i.e. $\Delta z = 0.04f$ for our cavity geometry, the frequency spacings of the transverse modes are extended to the entire free spectral range (middle panel). However, the degeneracy in quality factors is also lifted, and many modes suffer a severe quality factor degradation (right panel). Therefore, the number of lasing modes will be significantly reduced, resulting in an effectively higher speckle contrast. 

In order to break the frequency degeneracy and increase the frequency spacings of the transverse modes, while minimizing their quality factor degradation, we explore a different approach, i.e. inserting an intracavity phase diffuser into the DCL (Figure~\ref{fig:1_exp_skecth}C). The phase diffuser is placed inside the DCL, next to the output coupler, in order to maintain the self-imaging condition of the cavity. The intracavity phase diffuser is a computer-generated random phase plate made of glass. It introduces an optical phase delay that varies randomly from $-\pi$ to $\pi$ on a length scale of $\approx200$ \textmu m (see Methods (1)). The middle panel of Figure~\ref{fig:1_exp_skecth}C shows that the transverse modes are spread over the entire free spectral range (FSR) of the DCL, increasing the frequency spacings between them. However, in contrary to the misaligned cavity case, many transverse modes experience minor quality factor degradation. As a result, a large number of transverse modes are expected to lase in a wide spectrum of frequencies, accelerating the speckle suppression process.

\section{Ultrafast speckle suppression}

\begin{figure}[htbp]
	\centering
	\includegraphics[width=1\linewidth]{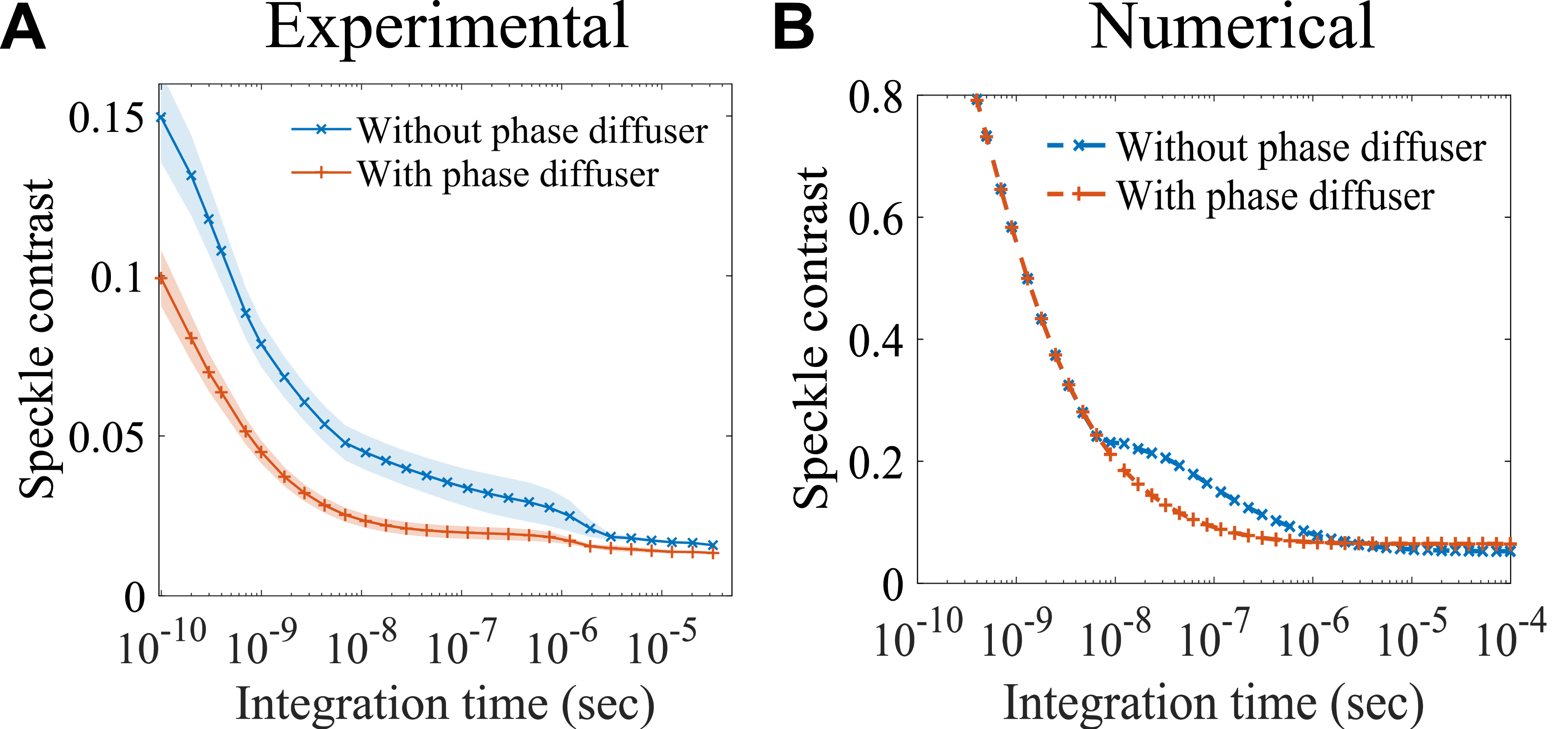}
	\caption{Speckle contrast as a function of the photodetector's integration time measured for the DCL without and with an intracavity phase diffuser. The pump power is roughly three times the lasing threshold ($P\approx3P_{\rm th}$). (A) Experimental data. (B) Numerical results. Experimentally, the intracavity phase diffuser introduces a significant reduction in speckle contrast for integration times $\tau$ less than $10^{-6}$ sec. Numerically, the reduction of speckle contrast by the intracavity phase diffuser is seen for $10^{-8}$ sec $< \tau < 10^{-6}$ sec. }
	\label{fig:2_C_vs_dt}
\end{figure}

To demonstrate the efficiency of our method, we experimentally measure the speckle contrast for integration times in the range of $10^{-10}$ to $10^{-4}$ sec. The output beam of the DCL is incident onto a thin diffuser placed outside the laser cavity. Then the speckle intensity is measured by a fast photodetector at the far field. See Methods (1) for a detailed description of the experimental setup and the measurement scheme. Figure~\ref{fig:2_C_vs_dt}A shows the measured speckle contrast as a function of the photodetector's integration time without and with an intracavity phase diffuser, at the pump power of 3 times the lasing threshold. By measuring the speckle contrast over many time windows of an equal length, we compute the mean contrast value and estimate the uncertainty that is shown by the shaded area. The lasing pulse is $\sim100$ \textmu s long. To avoid the transient oscillations at the beginning of the lasing pulse, we analyze the emission after the laser reaches a quasi steady state. For the effects of lasing transients, see Supplementary S1. Experimental data with a lower pump power is presented in Supplementary S2.

With the intracavity phase diffuser, the speckle contrast at short integration times (between $10^{-10}$ to $10^{-7}$ sec) is significantly lower than that without the intracavity phase diffuser. Even when the integration time is as short as $10^{-9}$ sec, the speckle contrast is already reduced to $3\%$. To understand this remarkable result, we numerically calculate the field evolution in a passive cavity with a simplified (1+1)D model. Nonlinear interactions of the lasing modes through the gain medium are neglected (see Methods (2) for details about the numerical model). The calculated speckle contrast is plotted as a function of integration time $\tau$ in Figure~\ref{fig:2_C_vs_dt}B. When $\tau$ is shorter than the inverse of the emission spectrum width $1/ \Delta \Omega$, all lasing modes within $ \Delta \Omega$ are mutually coherent with each other. The interference of their fields scattered by the external diffuser produces a speckle pattern of unity contrast ($C\approx 1$). 

Once $\tau > 1/ \Delta \Omega$, the lasing modes of frequency spacing larger than $1 / \tau$ decohere with respect to each other, and the intensity sum of their scattered light reduces the speckle contrast. With increasing $\tau$, more lasing modes become mutually incoherent, and the speckle contrast continues to drop. In a slightly imperfect DCL without the phase diffuser, the longitudinal mode spacing $\Delta \omega_l$ is much larger than the transverse mode spacing $\Delta \omega_t$. Once $\tau$ exceeds $1/ \Delta \omega_l \sim 10^{-8}$ sec, different longitudinal modal groups are mutually incoherent, but the transverse modes within each longitudinal modal group remain coherent till $\tau$ reaches $1/ \Delta \omega_t \sim 10^{-6}$ sec. Thus the speckle contrast reduction is greatly slowed down in the time interval between $10^{-8}$ and $10^{-6}$ sec. Once $\tau$ exceeds $10^{-6}$ sec, the decoherence of the transverse modes leads to a further reduction of speckle contrast. See Ref.~\citep{Chriki18} for a comprehensive theory. 

With an intracavity phase diffuser in the DCL, the gap between $\Delta \omega_l$ and $\Delta \omega_t$ diminishes as the intracavity phase diffuser introduces different phase delays (frequency shifts) to individual transverse modes (as depicted in Figure~\ref{fig:1_exp_skecth}C). Meanwhile, the intracavity phase diffuser causes a relatively small reduction in the quality factor of many transverse modes. Thus a large number of transverse modes can still lase and their frequency detuning accelerates the spatial decoherence. In the time interval of $10^{-8}$ to $10^{-6}$ sec, the speckle contrast continues to decrease due to the decoherence of the transverse modes within one free spectral range. 

\begin{figure}[htbp]
	\centering
	\includegraphics[width=1\linewidth]{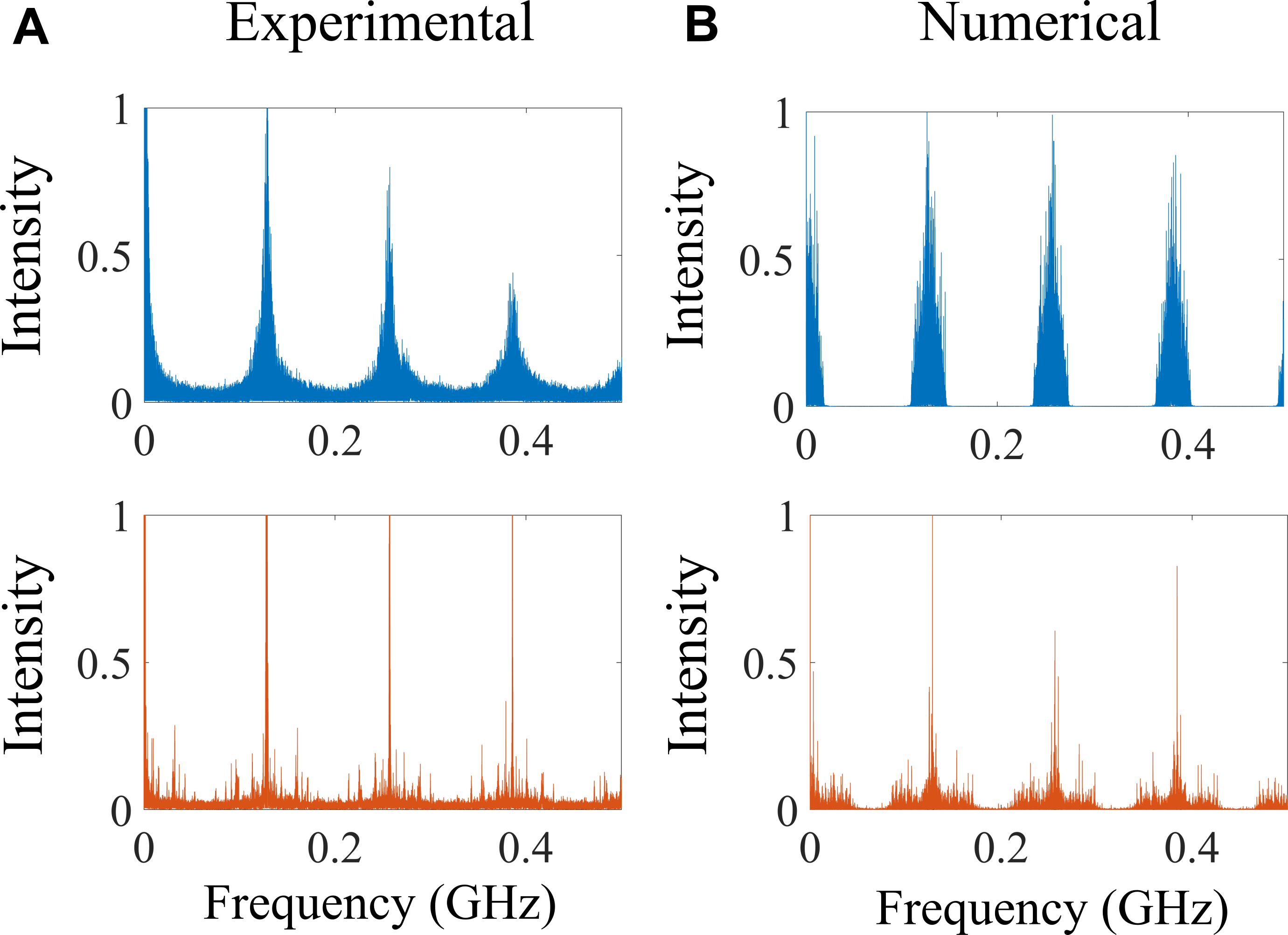}
	\caption{Power spectra of the DCL's emission intensity without (top row) and with (bottom row) the phase diffuser. (A) Experimental data. (B) Numerical results. The intracavity phase diffuser broadens the radio-frequency distribution in each FSR unit, increasing the frequency spacing of the transverse modes and leading to faster spatial decoherence and speckle suppression.}
	\label{fig:3_beat_f}
\end{figure}

To verify this explanation, we compare the power spectrum of emission intensity of the DCL with the intracavity phase diffuser to that without. The power spectrum spectrum is obtained by Fourier transforming the time intensity signal of the emission. Figure~\ref{fig:3_beat_f} shows the measured and simulated power spectra, which reflects the frequency beating of the lasing modes. Without the intracavity phase diffuser (top row), the power spectrum features narrow distributions peaked at the harmonics of FSR $= c/(2L)\approx128$ MHz, where $c$ is the speed of light, and $L = 117$ cm the total optical length of the DCL. The narrow distributions centered at the harmonics of the FSR reveal a slight breaking of frequency degeneracy of the transverse modes, due to the inherent imperfections of the cavity. 
With the intracavity phase diffuser (bottom row), the power spectrum of emission intensity features many narrow peaks in between the harmonics of the FSR. As the transverse modes move further away from the frequency degeneracy, their frequency differences, which determine their beat frequencies, increase. Nevertheless, the longitudinal mode spacing is unchanged, thus the peaks at the harmonics of the FSR remain in the power spectrum but appear narrower than that without the intracavity phase diffuser. The changes in the power spectrum indicates a frequency broadening of spatio-temporal modes by the intracavity phase diffuser. An ensemble of mutually incoherent lasing modes separated by frequency spacings in the range of $\sim 1$ MHz to $\sim128$ MHz leads to a faster decoherence rate on the time scale of $\sim 10^{-8}$ sec to $\sim 10^{-6}$ sec. This observation is consistent with the behavior shown in Figure~\ref{fig:2_C_vs_dt}.

\begin{figure}[htbp]
	\centering
	\includegraphics[width=1\linewidth]{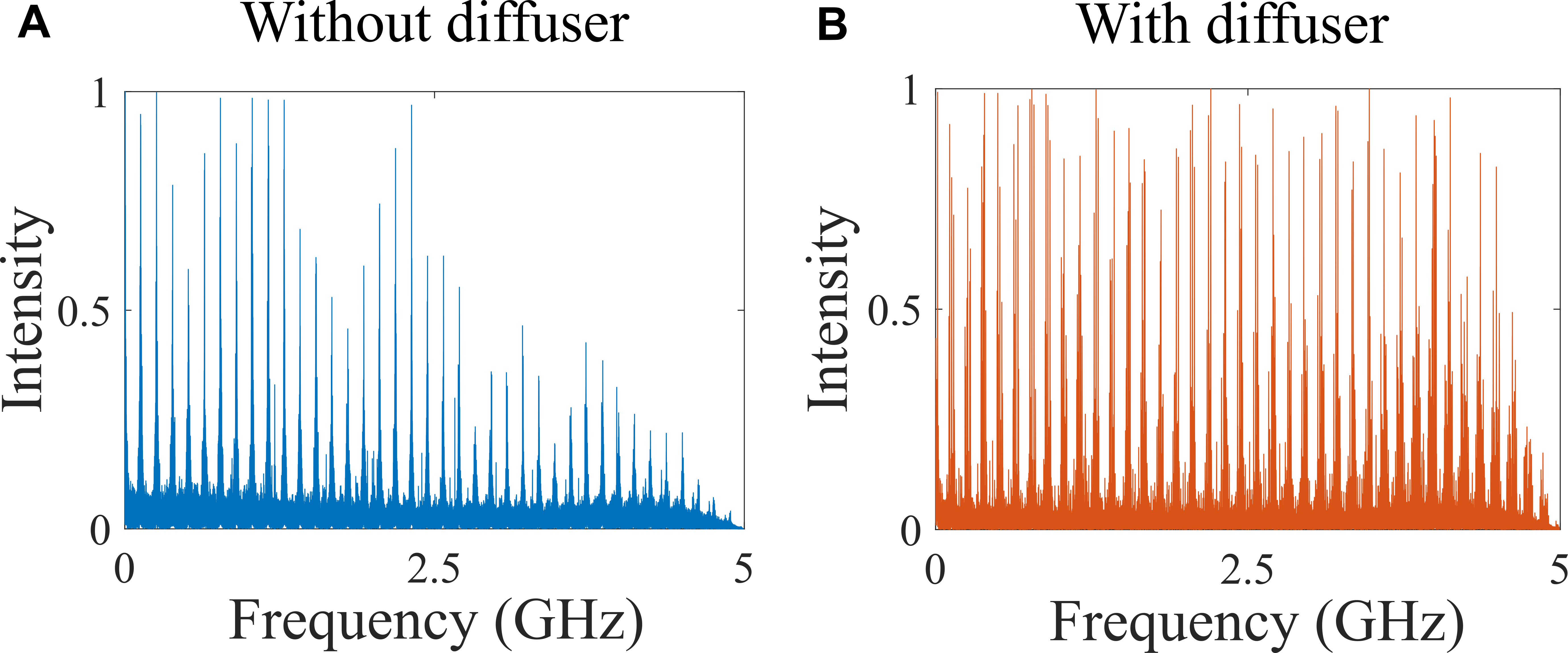}
	\caption{Experimentally-measured full-scale power spectrum of emission intensity of the degenerate cavity laser (A) without and (B) with the intracavity phase diffuser. The power spectrum envelope decreases with frequency in (A) and remains nearly constant in (B), indicating that the intracavity phase diffuser enhances lasing in a broader spectral range and accelerates speckle suppression within 1 nanosecond integration time.}
	\label{fig:4_full_beat_f_spectrum}
\end{figure}

Surprisingly, the intracavity phase diffuser causes a significant speckle contrast reduction even when the integration time is shorter than $10^{-8}$ sec, as seen in Figure~\ref{fig:2_C_vs_dt}A. Note that this behavior is not captured in the simulation (Figure~\ref{fig:2_C_vs_dt}B). To explain this effect, we analyze the entire experimentally-measured power spectra \citep{Note1}. The results are presented in Figure~\ref{fig:4_full_beat_f_spectrum} both (A) without and (B) with the phase diffuser in the DCL. 

Without the intracavity phase diffuser, the power spectrum envelope decays with increasing frequency. With the intracavity phase diffuser, the power spectrum exhibits an essentially constant envelope over the entire power detection range of $5$ GHz. This difference indicates that the intracavity phase diffuser facilitates lasing in a broader frequency range. With the intracavity phase diffuser, the mutually incoherent lasing modes of frequency spacing well above $1$ GHz accelerate the speckle reduction in the sub-nanosecond time scale. Our numerical simulation does not account for mode competition for gain, and cannot predict the lasing spectrum broadening. Therefore, the difference between the experimental and the numerical results in the ultrashort time regime is attributed to the change in nonlinear lasing dynamics. Namely, the intracavity phase diffuser reduces mode competition for gain, allowing modes with wider frequency differences to lase simultaneously.

Finally, we measure the total output power of the DCL without and with the intracavity phase diffuser. As shown in Supplementary S3, inserting the phase diffuser in the DCL results in a power loss of about $15\%$.

\section{Conclusion}

In conclusion, we introduce a novel method to accelerate the spatial decoherence of a degenerate cavity laser (DCL). The method relies on the insertion of an intracavity phase diffuser into a DCL to lift the frequency degeneracy of the transverse modes and broaden the lasing spectrum. Although the degeneracy of quality factors is also lifted, a large number of modes can still lase, due to relatively high quality factors. The frequency detuning of the modes enhances the speckle suppression at short integration times. In less than one nanosecond, the speckle contrast is already reduced to $3\%$. The laser power reduction by the intracavity phase diffuser is about $15 \%$. Such a fast spatial decoherence is useful for time-resolved full-field imaging of transient phenomena such as the dynamics of material processing~\cite{MermillodBlondin13} and tracking of moving targets~\cite{Chriki18, Nixon13}. We plan to extend this work by further investigating how the intracavity phase diffuser modifies the nonlinear modal interactions and the spatiotemporal dynamics of a DCL~\citep{Bittner18}.

\section*{Methods}
\subsection*{(1) Detailed experimental setup} \label{section:expsetup}
\begin{figure}[htbp]
	\centering
	\includegraphics[width=1\linewidth]{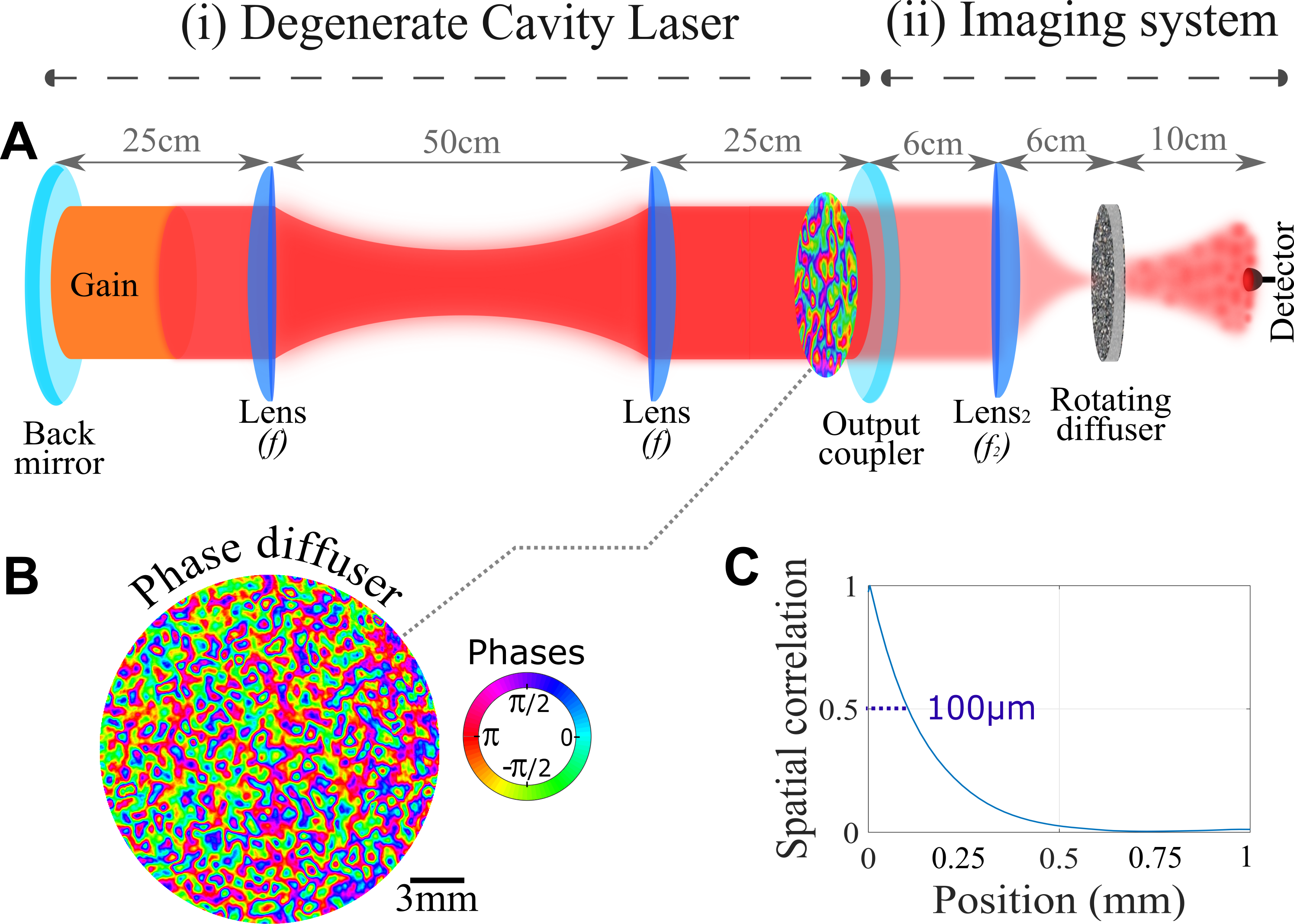}
	\caption{Experimental configuration for the time-resolved speckle intensity measurement. (A) Sketch of (i) a degenerate cavity laser (DCL) with an intracavity phase diffuser, and (ii) an imaging system with a external diffuser for generating speckle and a fast photodetector for measuring the speckle intensity as a function of time. (B) The two-dimensional phase profile of the intracavity phase diffuser, measured by a home-built optical interferometer. (C) Cross-section of the two-dimensional autocorrelation function of the phase profile shown in (B), its width gives the typical length scale over which the phase varies.}
	\label{fig:M1_detailed_exp_skecth}
\end{figure}

Our experimental setup, shown in Figure~\ref{fig:M1_detailed_exp_skecth}, consists of two parts: (i) a DCL with an intracavity phase diffuser, (ii) an imaging system to generate speckle with an external diffuser, and measure the speckle contrast~\citep{Chriki18}. The DCL is comprised of a flat back mirror with $95\%$ reflectivity, a Nd:YAG crystal rod of $10.9$ cm length and $0.95$ cm diameter, two spherical lenses of $5.08$ cm diameter and $f=25$ cm focal length, and an output coupler with $80\%$ reflectivity. Adjacent to the output coupler, the phase diffuser is placed inside the cavity. Lasing occurs at the wavelength of 1064 nm with optical pumping. The output beam is focused by a lens with a diameter of $2.54$ cm and a focal length of $f_{2}=6$ cm onto a thin diffuser with a $10^{\circ}$ angular spread of the transmitted light. A photodetector with a $0.1$ ns integration time and a $30$ \textmu m diameter is placed at a distance of $10$ cm from the diffuser and records the scattered light intensity within a single speckle grain in time. We rotate the diffuser and repeat the time intensity trace measurement of a different speckle grain. In total, $100$ time intensity traces are recorded. 

The intracavity phase diffuser (Figure~\ref{fig:M1_detailed_exp_skecth}B) is a computer-generated surface relief random phase plate of diameter $5.08$ cm and thickness $2.3$ mm. The angular spread of the transmitted light is $0.3^{\circ}$. The two-dimensional phase profile across the phase diffuser is measured with a home-built optical interferometer. The phase randomly varies in $16$ equal steps between $-\pi$ to $\pi$ with a uniform probability density. From the measured phase profile, we compute the spatial correlation function, as shown in Figure~\ref{fig:M1_detailed_exp_skecth}C. Its half width at half maximum is $100$ \textmu m. The spatial correlation length is $200$ \textmu m, in agreement with the angular spread of the scattered light from the phase diffuser. 

\subsection*{(2) Numerical simulation} \label{section:numericalsim}
We simulate continuous wave propagation in a passive degenerate cavity with and without the intracavity phase diffuser. The cavity length and width are identical to those of the DCL in our experiment, except that the cross-section is one dimensional in order to shorten the computation time. Without the intracavity phase diffuser, the field evolution matrix of a single round trip in the cavity is 
\begin{equation}
M^{wo} = M_{B} \cdot M_\epsilon \cdot M_{F}
\end{equation}
where $M_{F}$ is the field propagation matrix from the back mirror to the output coupler, and $M_{B}$ from the output coupler to the back mirror, $M_\epsilon$ represents a small axial misalignment of the DCL~\citep{Arnaud69_2}. 
With the intracavity phase diffuser placed next to the output coupler, 
the field evolution matrix of a single round trip becomes:
\begin{equation}	
M^{w} = M_{B} \cdot M_\epsilon \cdot M_{PD} \cdot M_{F} ,
\label{eq:1}
\end{equation}
where $M_{PD}$ represents the phase delay of the field induced by the phase diffuser for one round trip in the cavity. To construct $M_{PD}$ in our simulations, we use the spatial distribution of the phase delay taken from the measured profile in Figure~\ref{fig:M1_detailed_exp_skecth}B. 

The matrices $M^{wo}$ and $M^{w}$ are diagonalized to obtain the eigenmodes of the cavity without and with the intracavity phase diffuser. A subset of the eigenmodes have high quality factors (low losses). Hence, they have low lasing threshold, and correspond to the lasing modes. The total field in the cavity can be expressed as a sum of these modes: 

\begin{equation}
E(x,t)=\sum\limits_{m=-M}^{M}\sum\limits_{n=1}^{N} \alpha_{m,n}\psi_{n}(x){\rm e}^{{\rm i}[\omega_{m,n} t + \phi_{m,n}(t)]},
\end{equation}
where $\alpha_{m,n}$ and $\omega_{m,n}$ denote the amplitude and frequency of a mode with a longitudinal index $m$ and a transverse index $n$, $\psi_n(x)$ represents the transverse field profile for the $n$-th eigenmode. The phase $\phi_{m,n}(t)$ fluctuates randomly in time to simulate the spontaneous-emission-induced phase diffusion that leads to spectral broadening~\citep{Durrett19}. The total number of transverse modes is $N$ and the number of longitudinal modes is $2M$. 

The optical gain spectrum is approximated as a Lorentzian function centered at $\omega_0$ with a full-width-at-half-maximum (FWHM) of $32$ GHz. All lasing modes are within the gain spectrum and their frequencies can be written as $\omega_{m,n} = \omega_0 + m \, \Delta \omega_l +  \omega_n$, where $\Delta \omega_l$ is the longitudinal mode spacing (FSR), $m = \{-M,...+M\}$, $M = 120$, and  $\omega_n$ is the transverse mode frequency. The total number of time steps in the simulation of field evolution is $10^6$, each step has the duration of $0.1$ ns. 
The power spectrum is calculated by Fourier transforming the time trace of the intensity $|E(x,t)|^2$. 

To generate an intensity speckle, we simulate the field propagation from the output coupler of the degenerate cavity to the external diffuser, and then from the diffuser to the far field. The field intensity at the far field is used to compute the speckle contrast as a function of the integration time [see Methods (3)].

\subsection*{(3) Measurement of speckle contrast} \label{section:specklecontrast}
We use the experimental setup in Figure~\ref{fig:M1_detailed_exp_skecth}A to measure the time-resolved intensity of a single speckle grain behind a diffuser that is placed outside of the DCL. Using the detector with an integration time of $dt=0.1$ ns, we record the intensity as a function of time with and without the phase diffuser inside the DCL. The time trace of the intensity is recorded at 100 spatial locations $\vec{r}_i$ = ($x_i,y_i$), where $i=1...100$, by rotating the external diffuser by $3.6^{\circ}$ for each realization. From the $100$ intensity traces, we calculate the speckle contrast $C$ as a function of the integration time $\tau$. First, the total time window $T$ is divided into $J = T / \tau$ intervals.  For the $j$-th interval, the intensity is integrated in time: $I_{j}(\vec{r}_i,\tau) = \int_{j\tau}^{(j+1)\tau} I(\vec{r}_i,t) \, dt $, where $I(\vec{r}_i,t)$ is the time trace of intensity measured at location $\vec{r}_i$. Then, the speckle contrast is calculated for the integration time of $\tau$ for the $j$-th interval:

\begin{equation}
C_{j}(\tau) = \frac{\sigma_{j}(\tau)}{\mu_{j}(\tau) },
\end{equation}
where $\sigma_{j}(\tau)=\sqrt{\langle I^2_{j}(\vec{r}_i,\tau)\rangle_{i}-\langle I_{j}(\vec{r}_i,\tau)\rangle_{i}^{2}}$ is the standard deviation and $\mu_{j}(\tau)=\langle I_{j}(\vec{r}_i,\tau) \rangle _{i}$ is the mean intensity over $i=1...100$ spatial locations.
Finally, we compute the mean speckle contrast over all time intervals of length $\tau$: $C(\tau) = \langle C_j(\tau) \rangle _j$. The uncertainty of $C(\tau)$ is estimated from the standard deviation: $\sigma_C(\tau)=\sqrt{\langle C_{j}(
	\tau)^{2}\rangle_{j}-C(\tau)^{2}}$. Repeating this method, we compute the speckle contrast for different integration times $\tau$ in the range from $10^{-10}$ to $10^{-4}$ sec~\citep{Chriki18}. 

\subsection*{Acknowledgements}
The authors thank Arnaud Courvoisier and Ronen Chriki for their advice and help in the measurements. This work is partially funded by the US-Israel Binational Science Foundation (BSF) under grant no. 2015509. The work performed at Yale is supported partly by the US Air Force Office of Scientific Research under Grant No. FA 9550-20-1-0129, and we acknowledge the computational resources provided by the Yale High Performance Computing Cluster (Yale HPC). The research done at Weizmann is supported by the Israel Science Foundation.

\bibliographystyle{unsrt}
\bibliography{main}

\newpage 

\section*{Supplementary material}

\subsection*{(S1) Speckle contrast from the entire lasing pulse}
\begin{figure}[htbp]
\centering
\includegraphics[width=1\linewidth]{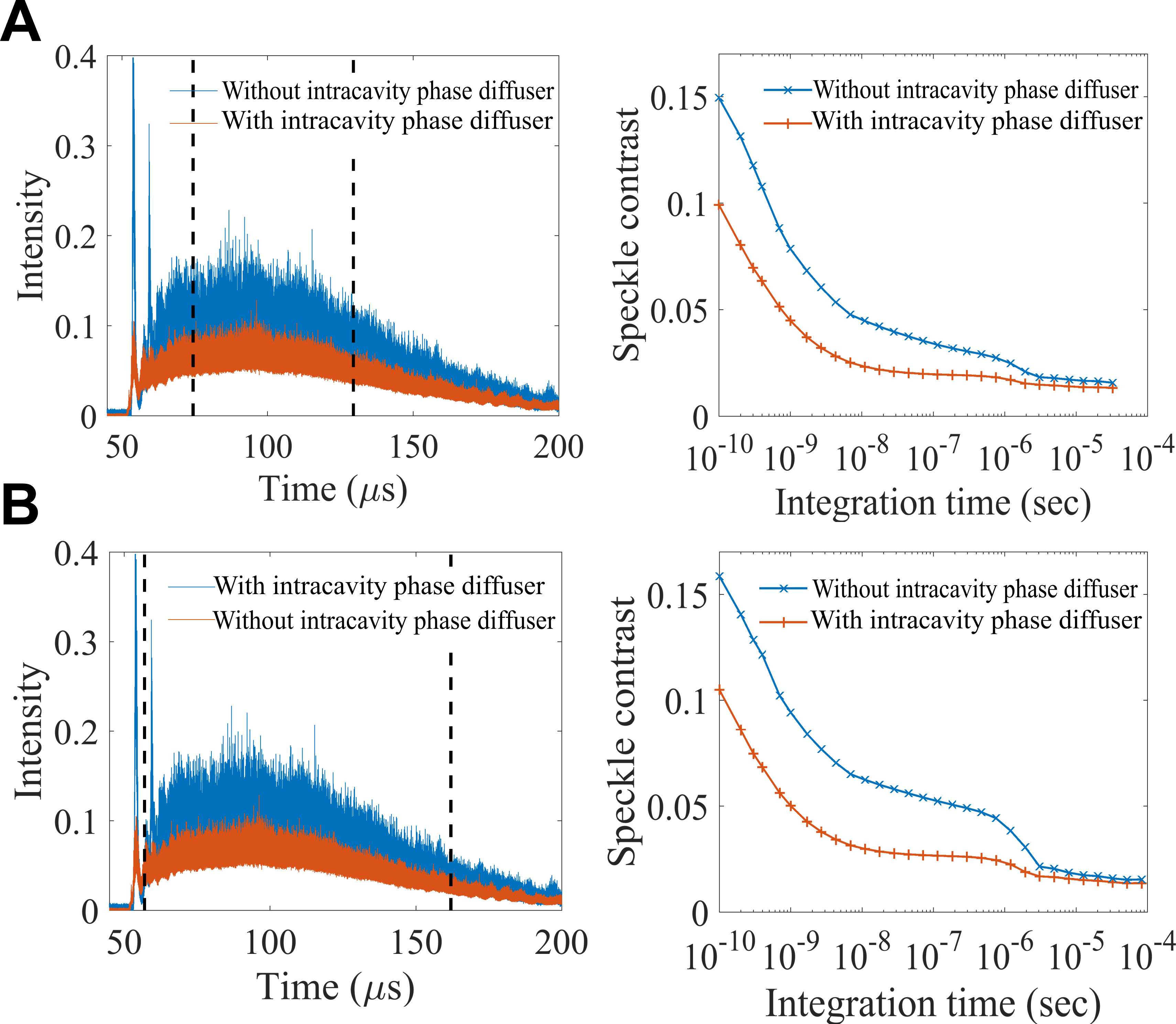}
\caption{Effects of lasing transients on speckle contrast. (A) The measured time trace of emission intensity (left panel) and the speckle contrast (right panel) obtained from the middle part (marked by vertical black dashed lines) of the lasing pulse where quasi steady state is reached. (B) The time intensity trace (left) and the speckle contrast (right) obtained from almost the entire lasing pulse that is marked by the vertical black dashed lines. When the lasing transients are excluded, the speckle contrast is lower, especially at integration times between $10^{-8}$ to $10^{-6}$ sec.}
\label{fig:S1_C_vs_t}
\end{figure}

For the speckle contrast analysis presented in Fig. 2 of the main text, we set the total time window $T$ to exclude any transient lasing behavior (e.g. spiking, relaxation oscillation) at the beginning of the lasing pulse and the intensity decay near the end of the lasing pulse. In Figure~\ref{fig:S1_C_vs_t}A, the vertical black dashed lines mark the time window of length $T \cong 50$ \textmu s, within which lasing reaches a quasi steady state, and the speckle contrast is obtained as a function of the integration time $\tau$.

Figure~\ref{fig:S1_C_vs_t}B shows the speckle contrast obtained with $T \cong 120$ \textmu s, including the lasing transients. The speckle contrast decreases with increasing $\tau$, similar to the case where the lasing transients are excluded. However, the speckle contrast is higher, especially at integration times between $10^{-8}$ to $10^{-6}$ sec, due to additional fluctuations of emission intensity at the beginning of the lasing pulse.

\newpage 

\subsection*{(S2) Speckle contrast at two different pump values}
\begin{figure}
	\centering
	\includegraphics[width=1\linewidth]{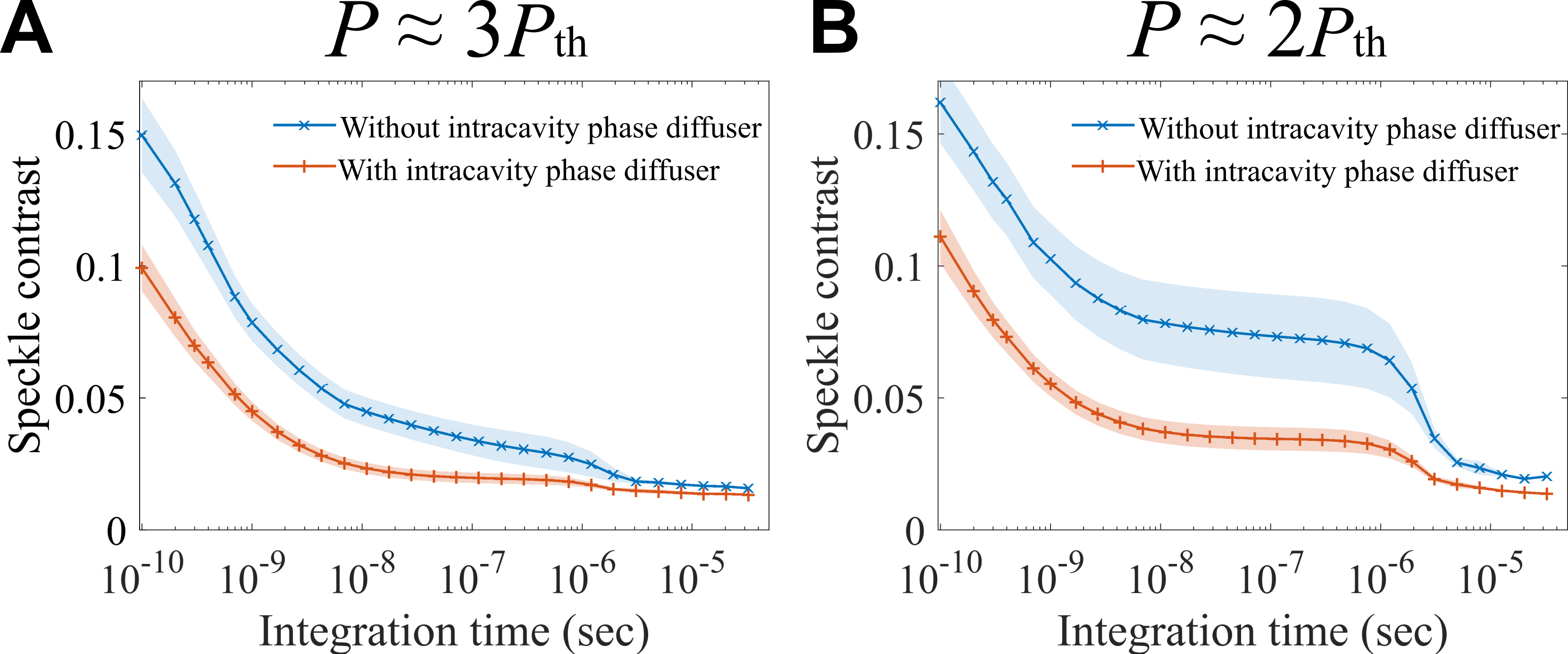}
	\caption{Speckle contrast as a function of the integration time, both without and with the intracavity phase diffuser in the DCL, at two pump levels. (A) $P\approx3P_{\rm th}$ (as in Fig. 2 in the main text).  (B) $P\approx2P_{\rm th}$. While the speckle contrast is higher at a lower pump level, the reduction of the speckle contrast by the intracavity phase diffuser is greater.}
	\label{fig:S2_C_vs_dt_P}
\end{figure}

We measure the speckle contrast at different pump levels of the DCL. In the main text (Figure 2A), we show the speckle contrast measured at three times of the lasing threshold $P\approx3P_{\rm th}$ (also shown in Figure~\ref{fig:S2_C_vs_dt_P}A). In Figure~\ref{fig:S2_C_vs_dt_P}B we present the data measured at a lower pump level $P\approx2P_{\rm th}$. Since less transverse modes lase at a lower pump level, the speckle contrast is higher at any integration time, with and without the intracavity phase diffuser. However, the speckle contrast reduction by the intracavity phase diffuser is greater at the lower pump level. Lasing transient processes such as temporal spiking and relaxation oscillation last for a longer time at the lower pump level and that results in an increase of the speckle contrast. The intracavity phase diffuser reduces the quality factor of most lasing modes, and shortens the transient process. Thus, within the time window for the speckle analysis, the effect of lasing transients is weaker. This fact is, at least partly, responsible for the greater reduction of speckle contrast at a lower pump level.   

In Figure~S\ref{fig:S2_C_vs_dt_P}B we present the data measured at a lower pump level $P\approx2P_{\rm th}$. Since less transverse modes lase at a lower pump level, the speckle contrast is higher at any integration time, with and without the intracavity phase diffuser. However, the speckle contrast reduction by the intracavity phase diffuser is greater at the lower pump level. Lasing transient processes such as temporal spiking and relaxation oscillation last for a longer time at the lower pump level and that results with an increase of the speckle contrast. The intracavity phase diffuser reduces the quality factor of most lasing modes, and shortens the transient process. Thus, within the time window for the speckle analysis, the effect of lasing transients is weaker. This fact is, at least partly, responsible for the greater reduction of speckle contrast at a lower pump level.  \\ 

\subsection*{(S3) Total output power of the DCL configurations}
\begin{figure}[H]
\centering
\includegraphics[width=1\linewidth]{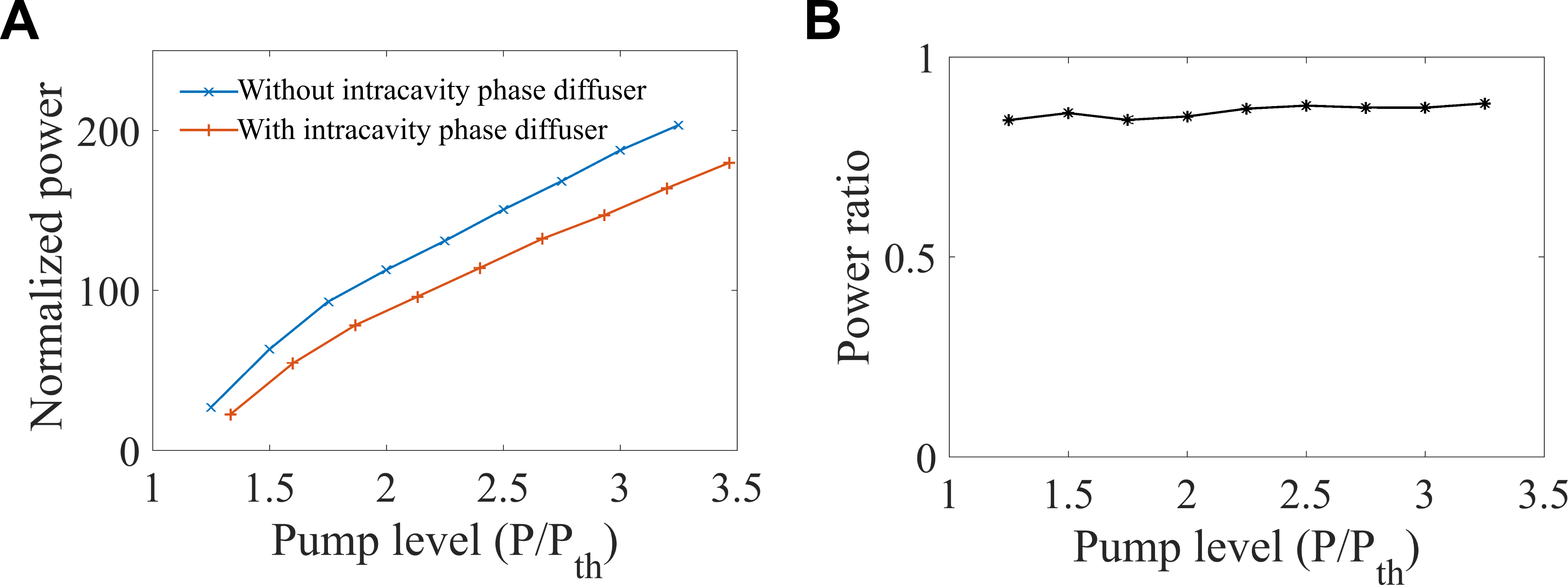}
\caption{Total output power of the degenerate cavity laser (DCL) with and without the intracavity phase diffuser. (A) The measured output power normalized by the value just above the lasing threshold as a function of the pump power normalized by the threshold value $P_{\rm th}$. (B) Ratio of the output power from the DCL with intracavity phase diffuser to that without. The intracavity phase diffuser reduces the output power by $15\%$ at all pump levels.}
\label{fig:S3_DCLs_total_power}
\end{figure}

We experimentally measure the total output power of the DCL without and with the intracavity phase diffuser. As shown in  Figure~\ref{fig:S3_DCLs_total_power}A, the total output power with the intracavity phase diffuser is slightly lower than that without. In Figure~\ref{fig:S3_DCLs_total_power}B we plot their ratio, which is about $0.85$ for all the pump levels. Thus, the intracavity phase diffuser causes a power reduction of about $15\%$. \\

\end{document}